# Phase trajectories of linear homogeneous autonomous dynamical systems of the third order

**A A Puntus[1] and A I Fedyushkin[2]**

[1] Moscow Aviation Institute, Volokolamsk highway, 4, Moscow, 125993, Russia
[2] Ishlinsky Institute for Problems in Mechanics of Russian Academy of Sciences, Prospekt Vernadskogo, 101-1, Moscow, 119526, Russia

E-mails: [1]artpuntus@yandex.ru, [2]fai@ipmnet.ru.

**Abstract**. The proposed article is devoted to the study of the problem of constructing phase trajectories in the vicinity of a singular point. This paper presents a more expanded view of this problem in comparison with those previously considered by other authors and suggests a more meaningful and grounded development of this study. Numerous examples of the application of this approach to the study of trajectories at a singular point are given.

## 1. Introduction

The paper is devoted to the study of the problem of constructing phase trajectories in the vicinity of a singular point - the origin of coordinates of a linear homogeneous autonomous dynamical system of the 3rd order. V.I. Arnold touched upon this problem in his textbook [1] in a rather limited volume, inviting the reader to continue this research. This work represents a more expanded view of this problem and offers a more meaningful and grounded development of this study, considered in [2].

The main content of this article is a well-founded study of the problem of constructing phase trajectories of a linear homogeneous autonomous dynamic system of the 3rd order in the vicinity of a singular point - the origin of coordinates, which, first of all, can be included in the curriculum of students and graduate students of universities [3, 4]. These results have not only theoretical significance for ordinary differential equations [6], but can be useful for studying the flow of liquid and gas, for example, used in the practical analysis of numerical results of CFD modeling.

## 2. Dynamical system of the 3rd order

Due to the complexity and novelty of this study, we first pay attention to the introduction of important definitions of invariant straight lines and invariant planes of linear homogeneous autonomous dynamical systems of the third order in the case of different roots of the characteristic equation of the system under study.

Consider a linear autonomous dynamical system of the 3rd order of the following type

$$\begin{pmatrix} \dot{x} \\ \dot{y} \\ \dot{z} \end{pmatrix} = \begin{pmatrix} a_{11} & a_{12} & a_{13} \\ a_{21} & a_{22} & a_{23} \\ a_{31} & a_{32} & a_{33} \end{pmatrix} \begin{pmatrix} x \\ y \\ z \end{pmatrix}$$

The matrix of this system $\mathbf{A} = \begin{pmatrix} a_{11} & a_{12} & a_{13} \\ a_{21} & a_{22} & a_{23} \\ a_{31} & a_{32} & a_{33} \end{pmatrix}$ can be considered as a linear operator transforming a vector $\begin{pmatrix} x \\ y \\ z \end{pmatrix}$ to the corresponding vector $\begin{pmatrix} \dot{x} \\ \dot{y} \\ \dot{z} \end{pmatrix}$. As is known from the course of linear algebra [5] if a linear operator $\mathbf{A}$ is given in a linear space that translates all vectors of a certain subspace into vectors of the same space, then this subspace is called invariant with respect to the operator $\mathbf{A}$.

Indeed, when the eigenvalues $\lambda_1$, $\lambda_2$, $\lambda_3$ – are real numbers, each eigenvector of a linear operator $\mathbf{A}$ corresponds to a one-dimensional invariant subspace, and each pair of eigenvectors belonging to different eigenvalues that are pairwise different from each other corresponds to a two-dimensional invariant subspace. Hence, if $\lambda_1$, $\lambda_2$, $\lambda_3$ – are real numbers, then in $R^3$ space there are three one-dimensional invariant subspaces (three invariant lines) and three two-dimensional invariant subspaces (three invariant planes). If the eigenvectors have coordinates

$$H_1 = \begin{pmatrix} x_1 \\ y_1 \\ z_1 \end{pmatrix}, \quad H_2 = \begin{pmatrix} x_2 \\ y_2 \\ z_2 \end{pmatrix}, \quad H_3 = \begin{pmatrix} x_3 \\ y_3 \\ z_3 \end{pmatrix},$$

then the canonical equations of invariant lines have the form:

$$\frac{x}{x_1} = \frac{y}{y_1} = \frac{z}{z_1}, \quad \frac{x}{x_2} = \frac{y}{y_2} = \frac{z}{z_2}, \quad \frac{x}{x_3} = \frac{y}{y_3} = \frac{z}{z_3}$$

Canonical equations of invariant planes containing invariant lines have the following forms:

$$\begin{vmatrix} x & x_1 & x_2 \\ y & y_1 & y_2 \\ z & z_1 & z_2 \end{vmatrix} = 0, \quad \begin{vmatrix} x & x_2 & x_3 \\ y & y_2 & y_3 \\ z & z_2 & z_3 \end{vmatrix} = 0, \quad \begin{vmatrix} x & x_3 & x_1 \\ y & y_3 & y_1 \\ z & z_3 & z_1 \end{vmatrix} = 0.$$

Due to the complexity and novelty of this study, we first pay attention to the introduction of important definitions of invariant straight lines and invariant planes of linear homogeneous autonomous dynamical systems of the third order with a constant matrix, that is the same:

$$\begin{pmatrix} \dot{x} \\ \dot{y} \\ \dot{z} \end{pmatrix} = \begin{pmatrix} a_{11} & a_{12} & a_{13} \\ a_{21} & a_{22} & a_{23} \\ a_{31} & a_{32} & a_{33} \end{pmatrix} \begin{pmatrix} x \\ y \\ z \end{pmatrix}.$$

The matrix of this system can be considered as a linear operator that transforms a vector $\begin{pmatrix} x \\ y \\ z \end{pmatrix}$ into the corresponding vector $\begin{pmatrix} \dot{x} \\ \dot{y} \\ \dot{z} \end{pmatrix}$.

Let 's make a characteristic equation of the matrix of this system: $\det(\mathbf{A} - \lambda E) = 0$ or else

$$\begin{vmatrix} a_{11} - \lambda & a_{12} & a_{13} \\ a_{21} & a_{22} - \lambda & a_{23} \\ a_{31} & a_{32} & a_{33} - \lambda \end{vmatrix} = 0.$$

At first, consider the case of real distinct roots of the characteristic equation of a given matrix. Revealing the determinant standing in the left part, we get: $\lambda^3 + \mu\lambda^2 + \beta\lambda + \gamma = 0$, where $\mu$, $\beta$, $\gamma$ – various real numbers, and $\mu = -(a_{11} + a_{22} + a_{33})$, $\gamma = -\det$.

Having solved the resulting cubic equation, we get the real roots $\lambda_1$, $\lambda_2$, $\lambda_3$. Since these roots are real and pairwise distinct, then by defining the eigenvectors $H_1$, $H_2$, $H_3$, we obtain the general solution of the system in the following form:

$$\begin{pmatrix} x \\ y \\ z \end{pmatrix} = C_1 H_1 e^{\lambda_1 t} + C_2 H_2 e^{\lambda_2 t} + C_3 H_3 e^{\lambda_3 t},$$

where vectors $H_1$, $H_2$, $H_3$ define invariant lines that are both eigenvectors and phase trajectories.

If $\lambda_1$ is a real root, and $\lambda_2$, $\lambda_3$ are complex-conjugate, and they are $\lambda_{2,3} = \mu \pm i\delta$, then the general solution of this system will have the form:

$$\begin{pmatrix} x \\ y \\ z \end{pmatrix} = C_1 H_1 e^{\lambda_1 t} + C_2 (H_2 \cos\delta t + H_3 \sin\delta t) e^{\mu t} + C_3 (H_2 \sin\delta t - H_3 \cos\delta t) e^{\mu t},$$

where $H_1$, $H_2$, $H_3$ are the vectors, which define the corresponding invariant straight lines and pairwise invariant planes.

Let's consider a concrete example of a linear autonomous dynamic system:

$$\begin{pmatrix} \dot{x} \\ \dot{y} \\ \dot{z} \end{pmatrix} = \begin{pmatrix} 3 & -3 & 1 \\ 3 & -2 & 2 \\ -1 & 2 & 0 \end{pmatrix} \begin{pmatrix} x \\ y \\ z \end{pmatrix}.$$

The characteristic equation of the matrix of this system $\begin{vmatrix} 3-\lambda & -3 & 1 \\ 3 & -2-\lambda & 2 \\ -1 & 2 & -\lambda \end{vmatrix} = 0$ has roots: $\lambda_1 = -1$, $\lambda_{2,3} = 1 \pm i$. The invariant line for $\lambda_1 = -1$ determines the solution of the system $\begin{cases} 4h_1 - 3h_2 + h_3 = 0 \\ 3h_1 - h_2 + 2h_3 = 0 \end{cases}$, i.e. the eigenvector $H_1 = \begin{pmatrix} 1 \\ 1 \\ -1 \end{pmatrix}$.

For complex roots $\lambda_{2,3} = 1 \pm i$, choosing a value $\lambda_2 = 1 + i$ and solving the system $\begin{cases} (2-i)h_1 - 3h_2 + h_3 = 0 \\ 3h_1 + (-3-i)h_2 + 2h_3 = 0 \end{cases}$, we determine the solution:

$$\begin{pmatrix} -1+i \\ i \\ 1 \end{pmatrix} e^{(1+i)t} = \left\{ \left[ \begin{pmatrix} -1 \\ 0 \\ 1 \end{pmatrix} \cos t + \begin{pmatrix} -1 \\ -1 \\ 0 \end{pmatrix} \sin t \right] + i \left[ \begin{pmatrix} -1 \\ 0 \\ 1 \end{pmatrix} \sin t - \begin{pmatrix} -1 \\ -1 \\ 0 \end{pmatrix} \cos t \right] e^t \right\}$$

Therefore, the general solution of this dynamical system has the form:

$$\begin{pmatrix} x \\ y \\ z \end{pmatrix} = C_1 \begin{pmatrix} 1 \\ 1 \\ -1 \end{pmatrix} e^{-t} + C_2 \left[ \begin{pmatrix} -1 \\ 0 \\ 1 \end{pmatrix} \cos t + \begin{pmatrix} -1 \\ -1 \\ 0 \end{pmatrix} \sin t \right] e^t + C_3 \left[ \begin{pmatrix} -1 \\ 0 \\ 1 \end{pmatrix} \sin t - \begin{pmatrix} -1 \\ -1 \\ 0 \end{pmatrix} \cos t \right] e^t.$$

The equations of invariant lines are: $\frac{x}{1}=\frac{y}{1}=\frac{z}{-1}$, $\frac{x}{-1}=\frac{y}{0}=\frac{z}{1}$ and $\frac{x}{-1}=\frac{y}{-1}=\frac{z}{0}$.

The equation of the invariant plane, including invariant lines corresponding to the complex conjugate roots of the characteristic equation, has the form: $\begin{vmatrix} x & -1 & -1 \\ y & 0 & -1 \\ z & 1 & 0 \end{vmatrix} = 0$ or $x+y+z=0$.

Let us now consider a linear autonomous dynamical system of the 3rd order of the following form

$$\begin{pmatrix} \dot{x} \\ \dot{y} \\ \dot{z} \end{pmatrix} = \begin{pmatrix} a_{11} & a_{12} & a_{13} \\ a_{21} & a_{22} & a_{23} \\ a_{31} & a_{32} & a_{33} \end{pmatrix} \begin{pmatrix} x \\ y \\ z \end{pmatrix}.$$

From a different point of view matrix $\mathbf{A} = \begin{pmatrix} a_{11} & a_{12} & a_{13} \\ a_{21} & a_{22} & a_{23} \\ a_{31} & a_{32} & a_{33} \end{pmatrix}$ of this system can be considered as a linear operator that transforms a vector $\begin{pmatrix} x \\ y \\ z \end{pmatrix}$ into the corresponding vector $\begin{pmatrix} \dot{x} \\ \dot{y} \\ \dot{z} \end{pmatrix}$. Linear systems of differential equations of this kind are considered in [6] in a similar operator form. As is known from the course of linear algebra [5] if a linear operator $\mathbf{A}$ is given in a linear space that translates all vectors of a certain subspace into vectors of the same space, then this subspace is called invariant with respect to the operator.

Indeed, when the roots $\lambda_1$, $\lambda_2$, $\lambda_3$ – are real numbers, each proper vector of a linear operator $\mathbf{A}$ corresponds to a one-dimensional invariant subspace, and each pair of eigenvectors belonging to different proper values that are pairwise different from each other corresponds to a two-dimensional invariant subspace. Consequently, if $\lambda_1$, $\lambda_2$, $\lambda_3$ – are real numbers, then in $\mathbf{R}^3$ there are three one-dimensional invariant subspaces (three invariant lines) and three two-dimensional invariant subspaces (three invariant planes). If the eigenvectors have coordinates are the same as

$$H_1 = \begin{pmatrix} x_1 \\ y_1 \\ z_1 \end{pmatrix}, \quad H_2 = \begin{pmatrix} x_2 \\ y_2 \\ z_2 \end{pmatrix}, \quad H_3 = \begin{pmatrix} x_3 \\ y_3 \\ z_3 \end{pmatrix},$$

then the canonical equations of invariant lines have the following forms:

$$\frac{x}{x_1}=\frac{y}{y_1}=\frac{z}{z_1}, \quad \frac{x}{x_2}=\frac{y}{y_2}=\frac{z}{z_2}, \quad \frac{x}{x_3}=\frac{y}{y_3}=\frac{z}{z_3}.$$

Canonical equations of invariant planes containing invariant lines have the form:

$$\begin{vmatrix} x & x_1 & x_2 \\ y & y_1 & y_2 \\ z & z_1 & z_2 \end{vmatrix} = 0, \quad \begin{vmatrix} x & x_2 & x_3 \\ y & y_2 & y_3 \\ z & z_2 & z_3 \end{vmatrix} = 0, \quad \begin{vmatrix} x & x_3 & x_1 \\ y & y_3 & y_1 \\ z & z_3 & z_1 \end{vmatrix} = 0.$$

Let us now turn to a possible variant when there is a real root $\lambda_1$ and a pair of complex conjugate roots $\lambda_2$ and $\lambda_3$ or $\lambda_{2,3} = \mu \pm i\delta$ among the roots of the characteristic equation of the matrix $\mathbf{A}$. In this case, for a given dynamical system, it is possible to specify the corresponding 3-dimensional invariant space, the coordinates of which are the invariant line corresponding to the real root and two alternative lines corresponding to complex conjugate roots. In this case, the general solution of this system will look like:

$$\begin{pmatrix} x \\ y \\ z \end{pmatrix} = C_1 H_1 e^{\lambda_1 t} + C_2 \left( H_2 \cos\delta t + H_3 \sin\delta t \right) e^{\mu t} + C_3 \left( H_2 \sin\delta t - H_3 \cos\delta t \right) e^{\mu t},$$

where are the vectors $H_1 = \begin{pmatrix} x_1 \\ y_1 \\ z_1 \end{pmatrix}$, $H_2 = \begin{pmatrix} x_2 \\ y_2 \\ z_2 \end{pmatrix}$, $H_3 = \begin{pmatrix} x_3 \\ y_3 \\ z_3 \end{pmatrix}$ invariant straight lines and pairwise invariant planes are defined. Consequently, this linear operator **A** has a unique one-dimensional and a unique two-dimensional invariant subspaces. For the eigenvalue $\lambda_1$ this one-dimensional invariant subspace is an invariant line whose canonical equation has the form $\dfrac{x}{x_1} = \dfrac{y}{y_1} = \dfrac{z}{z_1}$, and for complex conjugate vectors $\lambda_{2,3} = \mu \pm i\delta$ a two-dimensional invariant subspace is a plane, whose canonical equation has the form $\begin{vmatrix} x & x_2 & x_3 \\ y & y_2 & y_3 \\ z & z_2 & z_3 \end{vmatrix} = 0$.

These two spaces form a 3-dimensional invariant space whose coordinates are an invariant line corresponding to a real root and two invariant lines corresponding to complex conjugate roots.

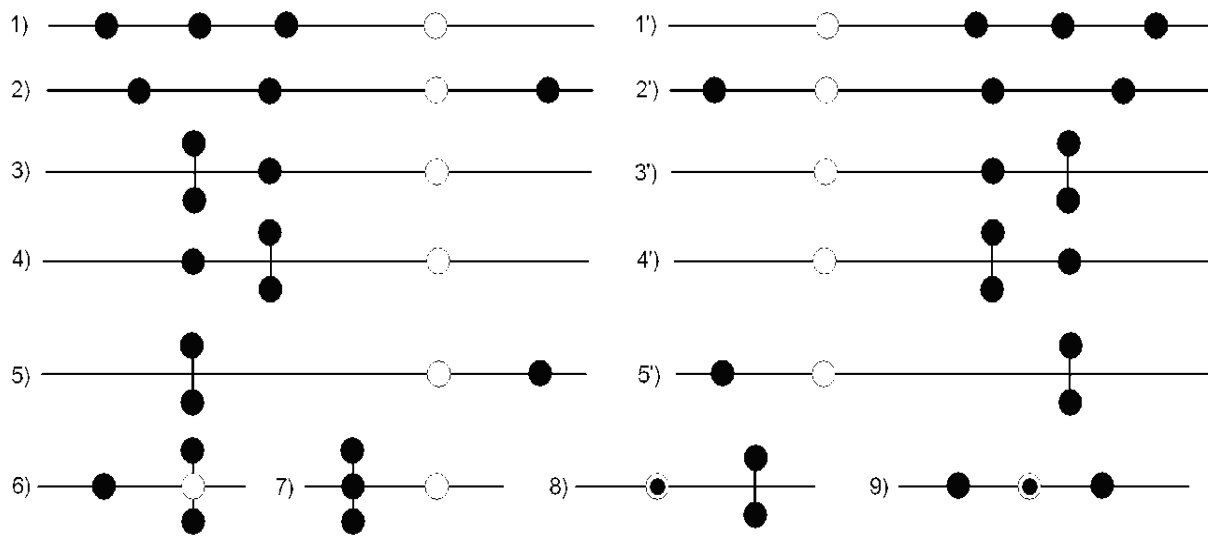

**Figure 1.** Possible location of the considered roots of the characteristic equation of the matrix **A** of a linear homogeneous autonomous dynamical system of the 3rd order.

Let us now consider concrete examples of constructing phase trajectories in the vicinity of a singular point - the origin of linear homogeneous autonomous dynamical systems of the 3rd order in the presence of different real roots of the characteristic equation of the matrix **A** or a real one and a pair of complex conjugate roots. All considered cases of such roots are represented by the following table, similar to [1], shown in figure1, possible visual conditional variants of the location of the roots relative to the origin and the real axis. In this case, the point denotes the value of the real root and two vertically arranged points – a pair of complex-conjugate roots relative to zero on the real axis. So, the possible variants of various real roots or real and pairs of complex-conjugate roots are as follows than are shown in figure1:

The corresponding image of the phase trajectories in the cases under consideration, shown in figure 1 has the form further presented in the following figure 2.

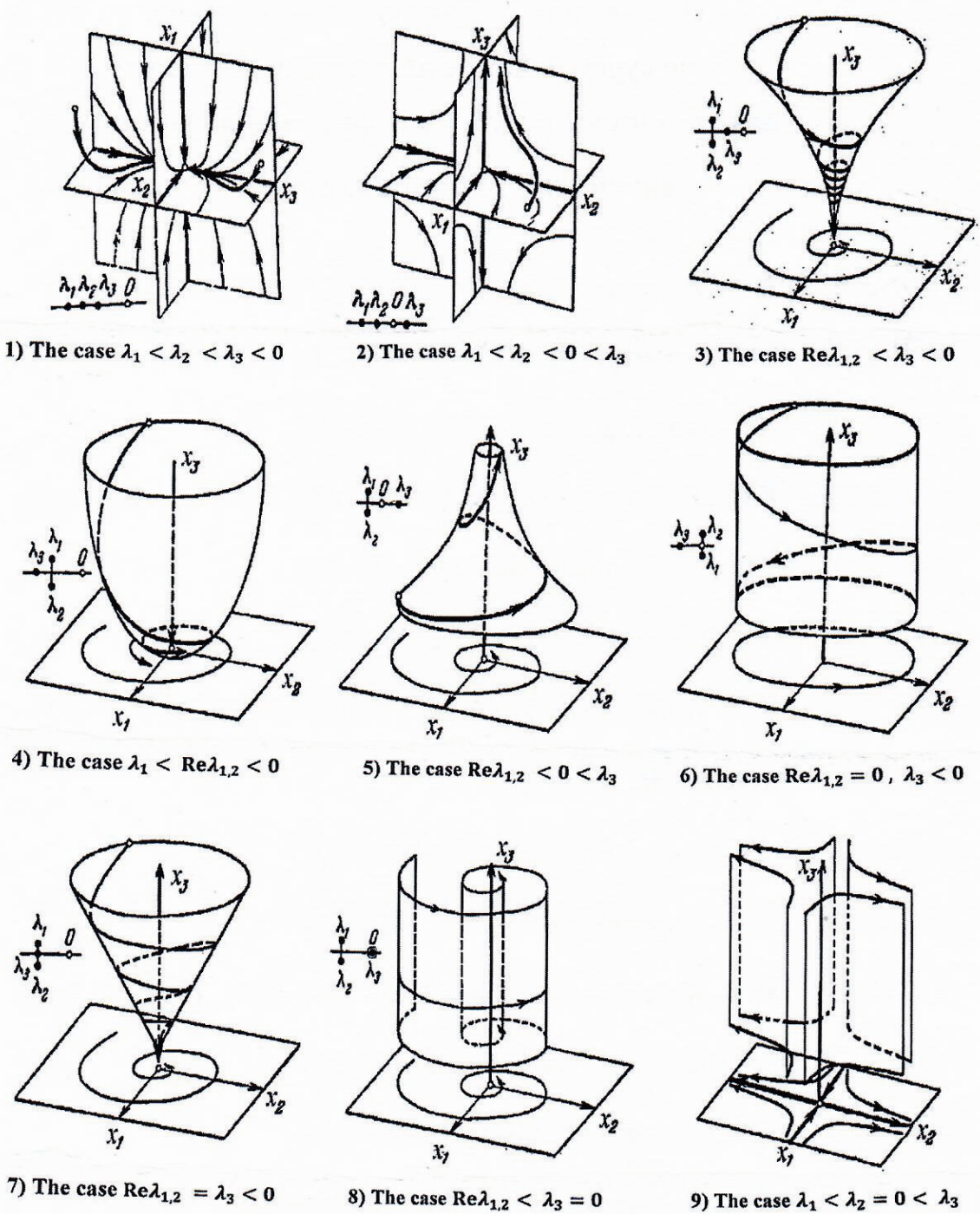

**Figure 2.** The picture of phase trajectories in the vicinity of the corresponding singular point of the 3rd order system in an orthogonal coordinate system.

Note that for clarity of the graphical representation of the phase trajectories of the considered linear homogeneous autonomous dynamical systems of the 3rd order near a special zero point, if the invariant lines are not orthogonal to each other, then it is necessary to orthogonalize the set of invariant lines and, consequently, transformation of a coordinate system $Oxyz$ into an orthogonal system $O\xi_1\xi_2\xi_3$ using one of the well-known from the algebra course [5] method of orthogonalization by Gram Schmidt or by transformation of the matrix A by a non-degenerate transformation using a matrix of invariant vectors to a Jordan structure from the corresponding orthogonal vectors (further, these orthogonal vectors define invariant lines - coordinate axes of the orthogonal system $O\xi_1\xi_2\xi_3$). We also assume in these examples, when illustrating the set of phase trajectories of the dynamical systems under consideration, that if such orthogonalization is required, then this transformation is implemented and this system of invariant straight lines and invariant planes is brought to their orthogonal location in three-dimensional space in figure 2.

Thus, when constructing a phase portrait of an autonomous dynamical system of the 3rd order, phase trajectories are already constructed taking into account the type of these trajectories on the invariant planes arranged orthogonally. In this case, these invariant lines are orthogonal coordinates.

For the noted uniformity of the figures, we also note that if the $Oxyz$ system obtained by solving a specific example turns out to be orthogonal, then it will be renamed in the figure to the $O\xi_1\xi_2\xi_3$ system by analogy with [1,2]. Thus, in all examples of graphical construction of phase trajectories near a singular point – the origin of coordinates of an autonomous dynamical system of the 3rd order, the corresponding system of invariant lines, invariant planes and phase trajectories is depicted in the orthogonal coordinate system $O\xi_1\xi_2\xi_3$.

The manual [2] provides specific examples illustrating the data presented in figures 1 and 2 for specific examples of an autonomous dynamical system of the 3rd order.

Next, we will consider some of them in this article as concrete examples of phase trajectories of an autonomous dynamical system of the 3rd order in the vicinity of the corresponding singular point of this system in the orthogonal coordinate system $O\xi_1\xi_2\xi_3$.

### 2.1. Example 1

An autonomous dynamic system is given: $\begin{cases} \dot{x} = 2x - y + z \\ \dot{y} = x + 2y - z \\ \dot{z} = x - y + 2z \end{cases}$.

*Decision.* The roots of the characteristic equation: $\lambda_1 = 1$, $\lambda_2 = 2$, $\lambda_3 = 3$.

Eigenvectors: $H_1 = \begin{pmatrix} 0 \\ 1 \\ 1 \end{pmatrix}$, $H_2 = \begin{pmatrix} 1 \\ 1 \\ 1 \end{pmatrix}$, $H_3 = \begin{pmatrix} 1 \\ 0 \\ 1 \end{pmatrix}$ define invariant lines $\frac{x}{0} = \frac{y}{1} = \frac{z}{1}$, $\frac{x}{1} = \frac{y}{1} = \frac{z}{1}$ and $\frac{x}{1} = \frac{y}{0} = \frac{z}{1}$, which correspond to the invariant vectors: $m_1 = (0,1,1)$, $m_2 = (1,1,1)$ and $m_3 = (1,0,1)$. These vectors are transformed using the orthogonal Gram Schmidt transformation [5] into an orthogonal system of vectors defining the orthogonal Cartesian coordinate system $O\xi_1\xi_2\xi_3$. In this coordinate system, the pattern of phase trajectories has the form (figure 3).

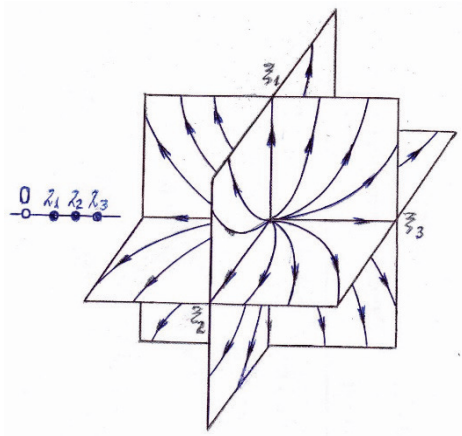

**Figure 3.** Phase portrait of the autonomous dynamic system of Example 1.

*2.2. Example 2*

An autonomous dynamic system is given $\begin{cases} \dot{x} = -x - 3y + z \\ \dot{y} = 2x - 3y + 2z \\ \dot{z} = 2x - y \end{cases}$

*Decision.* The roots of the characteristic equation: $\lambda_1 = -2$, $\lambda_{2,3} = -1 \pm 2i$.

The general solution of this system has the form:

$$\begin{pmatrix} x \\ y \\ z \end{pmatrix} = C_1 \begin{pmatrix} 1 \\ 0 \\ -1 \end{pmatrix} e^{-2t} + C_2 \left[ \begin{pmatrix} 0 \\ 1 \\ 1 \end{pmatrix} \cos 2t + \begin{pmatrix} -1 \\ 0 \\ 0 \end{pmatrix} \sin 2t \right] e^{-t} + C_3 \left[ \begin{pmatrix} 0 \\ 1 \\ 1 \end{pmatrix} \sin 2t - \begin{pmatrix} -1 \\ 0 \\ 0 \end{pmatrix} \cos 2t \right] e^{-t}.$$

Consequently, the equations of invariant lines have the form: $\dfrac{x}{1} = \dfrac{y}{0} = \dfrac{z}{-1}$, $\dfrac{x}{0} = \dfrac{y}{1} = \dfrac{z}{1}$ and $\dfrac{x}{-1} = \dfrac{y}{0} = \dfrac{z}{0}$. We transform these invariant lines into the corresponding system of orthogonal lines defining the orthogonal Cartesian coordinate system $O\xi_1\xi_2\xi_3$. In this coordinate system, the pattern of phase trajectories has the form (figure 4).

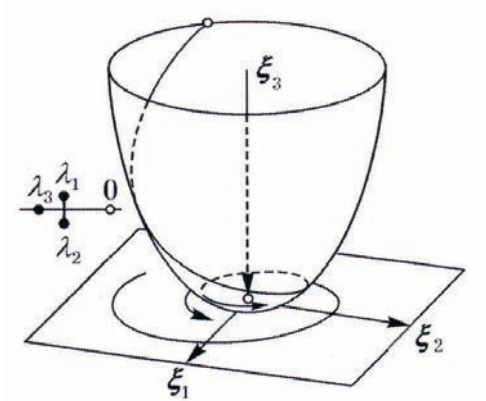

**Figure 4.** Phase portrait of the autonomous dynamic system of Example 2.

## 2.3. Example 3

An autonomous dynamic system is given: $\begin{cases} \dot{x} = -2x + 3y + 2z \\ \dot{y} = -3x + 3y + z \\ \dot{z} = 2x - y \end{cases}$

*Decision.* The roots of the characteristic equation: $\lambda_1 = -1$, $\lambda_{2,3} = 1 \pm i$.

The general solution of this system has the form:

$$\begin{pmatrix} x \\ y \\ z \end{pmatrix} = C_1 \begin{pmatrix} 1 \\ 1 \\ -1 \end{pmatrix} e^{-t} + C_2 \left[ \begin{pmatrix} 0 \\ -1 \\ 1 \end{pmatrix} \cos t + \begin{pmatrix} 1 \\ 1 \\ 0 \end{pmatrix} \sin t \right] e^t + C_3 \left[ \begin{pmatrix} 0 \\ -1 \\ 1 \end{pmatrix} \sin t - \begin{pmatrix} 1 \\ 1 \\ 0 \end{pmatrix} \cos t \right] e^t.$$

Consequently, the equations of invariant lines have the form: $\dfrac{x}{1} = \dfrac{y}{1} = \dfrac{z}{-1}$, $\dfrac{x}{0} = \dfrac{y}{-1} = \dfrac{z}{1}$ and $\dfrac{x}{1} = \dfrac{y}{1} = \dfrac{z}{0}$. We transform these invariant lines into the corresponding system of orthogonal lines defining the orthogonal Cartesian coordinate system $O\xi_1\xi_2\xi_3$. In this coordinate system, the pattern of phase trajectories has the form (figure 5).

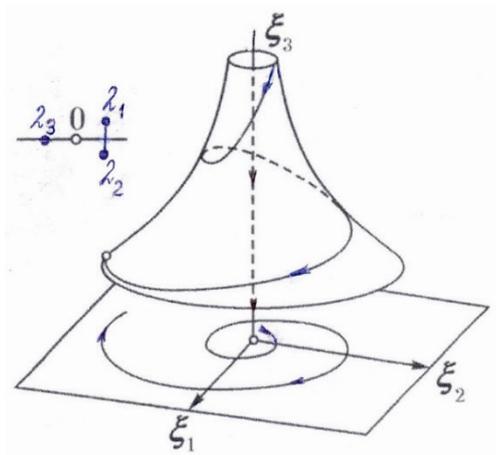

**Figure 5.** Phase portrait of the autonomous dynamic system of Example 3.

## 2.4. Example 4

An autonomous dynamic system is given: $\begin{cases} \dot{x} = 4x - 2y - 2z \\ \dot{y} = 3x - y - 2z \\ \dot{z} = 2x - y \end{cases}$

*Decision.* The roots of the characteristic equation: $\lambda_1 = 1$, $\lambda_{2,3} = 1 \pm i$.

The general solution of this autonomous dynamic system has the form:

$$\begin{pmatrix} x \\ y \\ z \end{pmatrix} = C_1 \begin{pmatrix} 0 \\ 1 \\ -1 \end{pmatrix} e^t + C_2 \left[ \begin{pmatrix} 1 \\ 1 \\ 1 \end{pmatrix} \cos t + \begin{pmatrix} 1 \\ 1 \\ 0 \end{pmatrix} \sin t \right] e^t + C_3 \left[ \begin{pmatrix} 1 \\ 1 \\ 1 \end{pmatrix} \sin t - \begin{pmatrix} 1 \\ 1 \\ 0 \end{pmatrix} \cos t \right] e^t.$$

Consequently, the equations of invariant lines have the form: $\dfrac{x}{0}=\dfrac{y}{1}=\dfrac{z}{-1}$, $\dfrac{x}{1}=\dfrac{y}{1}=\dfrac{z}{1}$ and $\dfrac{x}{1}=\dfrac{y}{1}=\dfrac{z}{0}$. We transform these invariant lines into the corresponding system of orthogonal lines defining the orthogonal Cartesian coordinate system $O\xi_1\xi_2\xi_3$. In this coordinate system, the pattern of phase trajectories has the form (figure 6).

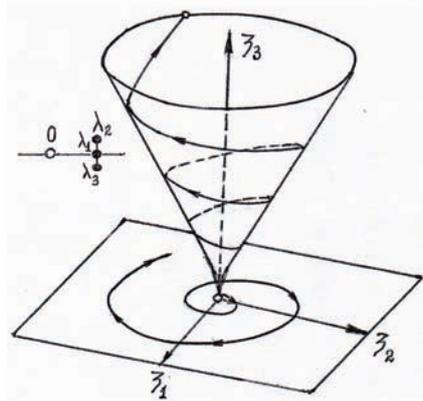

**Figure 6.** Phase portrait of the autonomous dynamic system of Example 4.

2.5. *Example 5*

An autonomous dynamic system is given: $\begin{cases} \dot{x}=2x-y+2z \\ \dot{y}=x+2z \\ \dot{z}=-2x+y-z \end{cases}$

*Decision.* The roots of the characteristic equation: $\lambda_1=1$, $\lambda_{2,3}=\pm i$.

The general solution of this autonomous dynamic system has the form:

$$\begin{pmatrix}x\\y\\z\end{pmatrix}=C_1\begin{pmatrix}0\\2\\1\end{pmatrix}e^t+C_2\left[\begin{pmatrix}1\\1\\1\end{pmatrix}\cos t+\begin{pmatrix}1\\1\\0\end{pmatrix}\sin t\right]+C_3\left[\begin{pmatrix}1\\1\\1\end{pmatrix}\sin t-\begin{pmatrix}1\\1\\0\end{pmatrix}\cos t\right].$$

Consequently, the equations of invariant lines have the form: $\dfrac{x}{0}=\dfrac{y}{2}=\dfrac{z}{1}$, $\dfrac{x}{1}=\dfrac{y}{1}=\dfrac{z}{1}$ and $\dfrac{x}{1}=\dfrac{y}{1}=\dfrac{z}{0}$. We transform these invariant lines into the corresponding system of orthogonal lines defining the orthogonal Cartesian coordinate system $O\xi_1\xi_2\xi_3$. In this coordinate system, the pattern of phase trajectories has the form (figure 7).

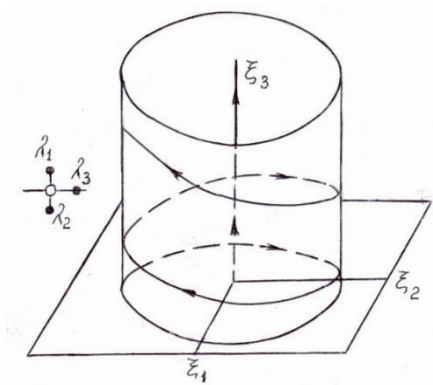

*2.6. Example 6*

An autonomous dynamic system is given: $\begin{cases} \dot{x} = -2x + y - z \\ \dot{y} = -x + y + z \\ \dot{z} = x - y - z \end{cases}$

*Decision.* The roots of the characteristic equation: $\lambda_1 = 0$, $\lambda_{2,3} = -1 \pm i$.

The general solution of this autonomous dynamic system has the form:

$$\begin{pmatrix} x \\ y \\ z \end{pmatrix} = C_1 \begin{pmatrix} 2 \\ 3 \\ -1 \end{pmatrix} + C_2 \left[ \begin{pmatrix} -1 \\ 1 \\ 1 \end{pmatrix} \cos t + \begin{pmatrix} 1 \\ 2 \\ 0 \end{pmatrix} \sin t \right] e^{-t} + C_3 \left[ \begin{pmatrix} -1 \\ 1 \\ 1 \end{pmatrix} \sin t - \begin{pmatrix} 1 \\ 2 \\ 0 \end{pmatrix} \cos t \right] e^{-t}.$$

Consequently, the equations of invariant lines have the form: $\dfrac{x}{2} = \dfrac{y}{3} = \dfrac{z}{-1}$, $\dfrac{x}{-1} = \dfrac{y}{1} = \dfrac{z}{1}$ and $\dfrac{x}{1} = \dfrac{y}{2} = \dfrac{z}{0}$. We transform these invariant lines into the corresponding system of orthogonal lines defining the orthogonal Cartesian coordinate system $O\xi_1\xi_2\xi_3$. In this coordinate system, the pattern of phase trajectories has the form (figure 8).

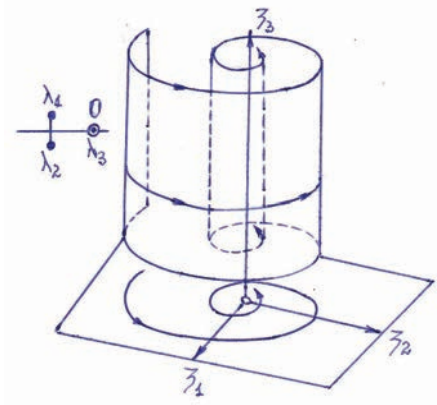

**Figure 8.** Phase portrait of the autonomous dynamic system of Example 6.

*2.7. Example 7*

An autonomous dynamic system is given: $\begin{cases} \dot{x} = -2x + y - z \\ \dot{y} = -x + y + z \\ \dot{z} = x - y - z \end{cases}$.

*Decision.* The roots of the characteristic equation: $\lambda_1 = 0$, $\lambda_2 = 1$, $\lambda_3 = -1$.

The general solution of this autonomous dynamic system has the form:

$$\begin{pmatrix} x \\ y \\ z \end{pmatrix} = C_1 \begin{pmatrix} 1 \\ 1 \\ 1 \end{pmatrix} + C_2 \begin{pmatrix} 1 \\ 0 \\ 1 \end{pmatrix} e^t + C_2 \begin{pmatrix} 1 \\ 1 \\ 2 \end{pmatrix} e^{-t}.$$

Consequently, the equations of invariant lines have the form: $\frac{x}{1}=\frac{y}{1}=\frac{z}{1}$, $\frac{x}{-1}=\frac{y}{0}=\frac{z}{1}$ and $\frac{x}{1}=\frac{y}{1}=\frac{z}{2}$. We transform these invariant lines into the corresponding system of orthogonal lines defining the orthogonal Cartesian coordinate system $O\xi_1\xi_2\xi_3$. In this coordinate system, the pattern of phase trajectories has the form (figure 9).

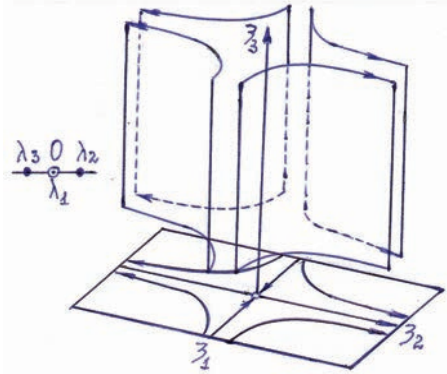

**Figure 9.** Phase portrait of the autonomous dynamic system of Example 7.

Similarly, examples for the remaining cases of the roots of the characteristic equation of the autonomous dynamical systems under consideration can be considered.

## 3. Conclusions
This article describes a method for studying phase trajectories in the vicinity of a singular point - the origin of coordinates of a linear homogeneous autonomous dynamical system of the 3rd order. The results of the application of this method of studying phase trajectories in the vicinity of a singular point and in other cases of the roots of the characteristic equation of the autonomous dynamical systems under consideration are presented on concrete examples. This research approach offers a more extended development and application of its results in the process of professional study and practical activity.